\documentclass[epj]{svjour}
\usepackage{graphics}
\begin{document}
\title{Chiral approach to antikaons in dense matter}
\author{Laura Tol\'os\inst{1} \and Angels Ramos\inst{2} \and Eulogio Oset\inst{3}%
}                     
%
%
\institute{Gesellschaft f\"ur Schwerionenforschung,
   Planckstrasse 1,
   D-64291 Darmstadt, Germany \and  Departament d'Estructura i Constituents de la Mat\`{e}ria,
     Universitat de Barcelona,
     Diagonal 647, 08028 Barcelona, Spain \and Departamento de F\'{\i}sica Te\'orica and IFIC, 
     Centro Mixto Universidad de Valencia-CSIC,
     Institutos de Investigaci\'on de Paterna, 
     Ap. Correos 22085, E-46071 Valencia, Spain }
\date{Received: date / Revised version: date}
%
\abstract{ Antikaons in dense nuclear matter are studied using a
chiral unitary approach which incorporates the $s$- and $p$-waves of the $\bar K N$ interaction. We 
include, in a self-consistent way, Pauli blocking effects, meson self-energies modified by nuclear 
short-range correlations and baryon binding potentials. We show that the on-shell factorization cannot be applied to evaluate the in-medium corrections to $p$-wave amplitudes. We also obtain an attractive shift for the $\Lambda$ and $\Sigma$ masses of -30 MeV at saturation density  while
the $\Sigma^*$ width gets sensibly increased to about 80 MeV.
The moderate attraction developed by the antikaon  does not support the existence of 
very deep and narrow bound states.
} 
\maketitle
Understanding the interaction of $\bar{K}$ with nucleons
and nuclei has recently been a matter of interest \cite{review}. The $\bar K N$ interaction below threshold is governed by the presence of the $\Lambda(1405)$ resonance. Phenomenology of kaonic atoms indicate that the $\bar K$ feels an attractive potential at low densities. Theoretically, this attraction can be traced back to the modified $\Lambda(1405)$ in the medium due to Pauli blocking effects \cite{koch} combined with the self-consistent consideration of the $\bar K$ self-energy \cite{lutz,schaffner} and the inclusion of self-energies of the mesons and baryons in the intermediate states \cite{Ramos:1999ku}. Moderate attractions of the order of -50 MeV at normal nuclear matter density $\rho_0$ are found by different approaches, in particular using chiral theories and the unitarization extensions in coupled-channels \cite{Ramos:1999ku}. 

Further theoretical advances have been achieved by looking at the $p$-wave contribution to the $\bar K N$ optical potential. While the $p$-wave contributions for atoms are negligible \cite{angelscarmen}, heavy-ion collisions provide an excellent scenario for testing high-momenta kaons and, therefore, further partial-wave contributions. In fact, some work has been done in that direction by studying the $\bar{K}$ at finite momenta from a self-consistent calculation using the J\"ulich meson exchange interaction \cite{laura}. An evaluation of antikaons in matter based on chiral dynamics including $s$-, $p$-, $d$-waves has been also performed \cite{lutz-korpa02}.

This work investigates the $\bar K$ in the nuclear medium following a chiral unitary approach using $s$- and $p$-wave contributions to $\bar K N$ and incorporating the corresponding many-body corrections. We will show that the on-shell factorization cannot be applied for $p$-waves in the medium as well as the importance of including the self-energy of mesons modified by nuclear short-range correlations and the  binding energy of baryons. We will obtain the $\bar K$ self-energy in nuclear matter and, as a byproduct, the self-energy of $\Lambda(1115)$, $\Lambda(1405)$, $\Sigma(1195)$ and $\Sigma^*(1385)$ \cite{tolos}.

\vspace{-0.5cm}
\section{In-medium $\bar K N$ interaction}

The properties of the $\bar K$ in the nuclear medium are obtained by incorporating medium modifications to the effective $\bar K N$  interaction. The effective $\bar K N$ interaction is built solving the coupled-channel Bethe-Salpeter equation using tree level contributions as the kernel of the equation. The kernel for the $s$-wave amplitude
is derived from the lowest order chiral lagrangian that couples the octet of pseudoscalar mesons to the octet of $1/2^+$ baryons \cite{oset}. The main contribution to the $p$-wave amplitudes comes from the $\Lambda$, $\Sigma$ and $\Sigma^*$ pole terms \cite{jido}. For meson-baryon scattering the kernel can be factorized on the mass shell in the loop functions \cite{oset,Oller:2000fj}. The loop function is then regularized by means of a cutoff or dimensional regularization. The formal result is schematically given by
\begin{equation}
    T = [1-VG]^{-1} V \label{BSeq} \ , 
\label{eq:tmatrix}
\end{equation}
where $V$ is the kernel and $G$ the loop function. 

The medium modifications arise from the inclusion of Pauli blocking effects on the nucleons
and the dressing of mesons and baryons in the intermediate loops. The binding effects for baryons are taken within the mean-field approach \cite{tolos}. For $\bar K$ and pions the medium modifications are included via the corresponding self-energy \cite{Ramos:1999ku}.


\begin{figure}
\resizebox{0.45\textwidth}{3cm}{
  \includegraphics{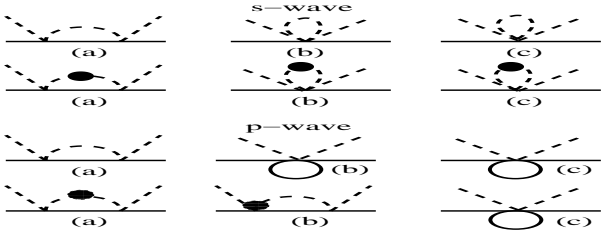}
}
\caption{On-shell (a), off-shell (b) and tadpole (c) contributions for $s$- (upper two rows) and $p$-wave (lower two rows) in free space and with self-energy insertions.}
\label{fig:1}       
\end{figure}

In order to calculate the in-medium amplitudes we will perform a similar unitarization procedure as in free space. The evaluation of the free space amplitudes rely on the factorization of the on-shell interaction kernel out of the loop function. In this section we show that the on-shell factorization is still valid for in-medium $s$-wave amplitudes while not for $p$-wave contributions.

In the $s$-wave amplitude of Fig.~\ref{fig:1}, the off-shell dependence of the two vertices in the loop function eliminates  a baryon propagator. Hence, this off-shell term ((b) in row 1) is cancelled by the presence of a tadpole term ((c) in row 1), in a suitable renormalization scheme. When we make self-energy insertions in the meson line, the cancellation between the off-shell part ((b) in row 2) and the tadpole ((c) in row 2) still holds. Then, the in-medium $s$-wave amplitudes are obtained by solving Eq.~(\ref{eq:tmatrix}) with the loop functions modified by meson self-energies and baryon binding corrections. 

The situation is different for the $p$-wave amplitudes due to the different $\vec{q}\,^2$ dependence (see Fig.~\ref{fig:1}). The off-shell part ((b) in row 3) eliminates a meson propagator (see Ref.~\cite{Cabrera:2002hc}) and, as a consequence, it can be cancelled by the tadpole term ((c) in row 3).  However, when the meson propagator is dressed, the off-shell term cancels only one of the two intermediate meson propagators. Hence, in the medium, we do not find the cancellation between the off-shell part ((b) in row 4) and tadpole term ((c) in row 4) that we found before for the $s$-wave amplitude. The use of the previous on-shell factorization could then lead to a violation of causality. This is solved by adding to the free loop the medium corrections calculated
using the full off-shell $\vec{q}\,^2$ contribution of the vertices
\vspace{-0.2cm}
\begin{eqnarray}
\label{gtogmedio}
&&G_l^p(s) \to G_l^p(s) + \frac{1}{\vec{q}\,^2_{on}} \lbrack I_{\rm med}(s) -
I_{\rm free}(s)
\rbrack \ ,\nonumber\\
&&I_{\rm med}(s) = i \int \frac{d^{4}q}{(2\pi)^{4}} \vec{q}\,^2 D_M(q) G_B(P-q)
\nonumber\\
&&I_{\rm free}(s) = i \int \frac{d^{4}q}{(2\pi)^{4}} \vec{q}\,^2 D^0_{M}(q)
G^0_{B}(P-q) \ .
\end{eqnarray}
with $D_M$ and $G_B$ being the meson and baryon propagators. The total four-momentum in the lab frame is  $P=q+p$, where $q$ and $p$ are the meson and baryon four-momentum in this frame. Another ingredient to be considered when dealing with $p$-wave amplitudes in the medium is the inclusion of nuclear short-range correlations. The $\pi$($\bar K$) propagators should account for the fact that the nucleon-nucleon (hyperon-nucleon) interaction is not only driven by one-pion (one-kaon) exchange (see details in Ref.~\cite{tolos}). 

The in-medium $p$-wave amplitudes are finally obtained from Eq.~(\ref{eq:tmatrix}) using the in-medium meson-baryon propagators of Eq.~(\ref{gtogmedio}), which incorporate the right $\vec{q}\,^2$ dependence, as well as Pauli blocking effects, dressing of mesons, baryon binding potentials and short-range correlations.

The $\bar K$ self-energy in nuclear matter is obtained self-consistently summing the in-medium $\bar K N$ for $s$- and $p$-waves over the Fermi sea of nucleons $n(\vec{p})$ according to
\begin{equation}
\Pi_{\bar{K}}(q^0,{\vec q},\rho)=4\int \frac{d^3p}{(2\pi)^3}\,
n(\vec{p}\,) \,  T_{\bar K N}(P^0,\vec{P},\rho) \ .
\label{eq:selfka}
\end{equation}
Then, the $\bar K$ spectral function reads
\begin{equation}
S_{\bar K}(q^0,{\vec
q},\rho)= -\frac{1}{\pi}\frac{{\rm Im} \Pi_{\bar K}(q^0,\vec{q},\rho)}
{\mid (q^0)^2-\vec{q}\,^2-m_{\bar K}^2-
\Pi_{\bar K} (q^0,\vec{q},\rho) \mid^2} \ .
\label{eq:spec}
\end{equation}

\begin{figure}
\resizebox{0.45\textwidth}{6cm}{
  \includegraphics{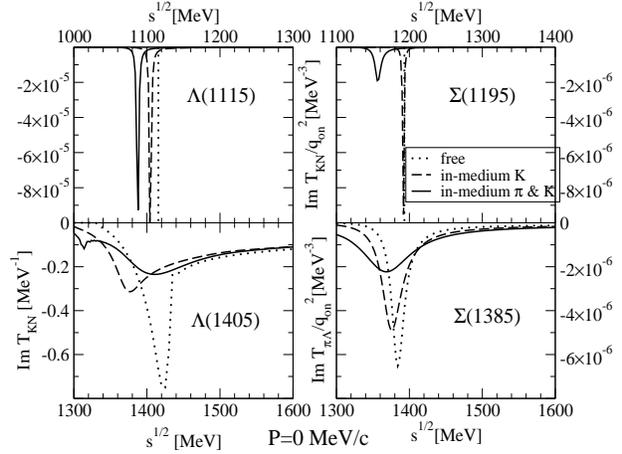}
}
\vspace{0.05cm}       
\caption{$\Lambda(1115)$, $\Lambda(1405)$, $\Sigma(1195)$ and $\Sigma^*(1385)$ resonances.}
\label{fig:2}       
\end{figure}
\vspace{-0.4cm}

\begin{figure*}
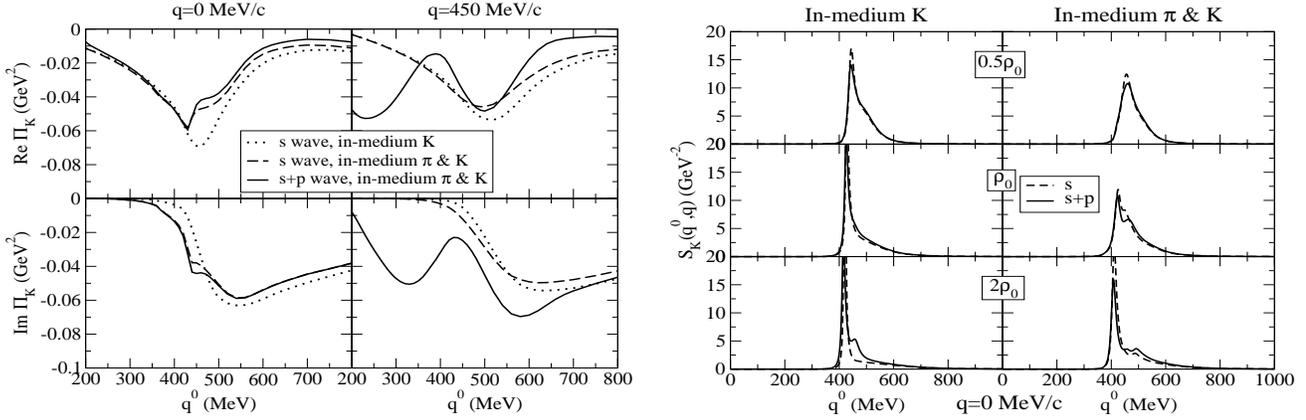

\vspace{0.13cm}
\begin{minipage}{200mm}
\resizebox{0.42\textwidth}{5.5cm}{
  \includegraphics{self.eps}}~~~
\resizebox{0.42\textwidth}{5.5cm}{
  \includegraphics{spec.eps}
}
\end{minipage}
\vspace{0.2cm}
\caption{$\bar K$ self-energy as function of $\bar K$ energy for two different kaon momenta (left). $\bar K$ spectral function  as function of $\bar K$ energy for zero momentum for the two approaches considered in the text (right). }
\label{fig:3}       
\end{figure*}

\vspace{-0.5cm}

\section{Hyperons and $\bar K$ self-energy}

In Fig.~\ref{fig:2} we display the results for the $\Lambda(1115)$, $\Lambda(1405)$, $\Sigma(1195)$ and $\Sigma^*(1385)$ by showing the imaginary part of the in-medium effective amplitude as function of c.m. energy $\sqrt{s}$ for a total momentum $P=0$. The free amplitudes (dotted lines) are compared to the in-medium ones at $\rho_0$ dressing the antikaons self-consistently (dashed lines) and also considering  the in-medium effects on pions (solid lines). The $\Lambda(1115)$ acquires an attractive shift of -10 MeV for the first approach while the shift is increased to -28 MeV when pions are dressed due to the appearance of new decay channels, namely $\Lambda N \to \Sigma N$. This is in accordance to hypernuclear spectroscopy data \cite{hyper}. The $\Lambda(1405)$ is generated dynamically close to the free position and gets strongly diluted when the in-medium properties of pions are considered as a consequence of new channels ($\Lambda N N^{-1}$, $\Sigma N N^{-1}$). The cusp seen for 1320 MeV corresponds to the opening of the $\pi \Sigma$ channel. With regards to the $I=1$ resonances, the  $\Sigma(1195)$ shows an attraction of -35 MeV when pions are dressed in line with \cite{Batty:1978sb} and in contrast with -10 MeV \cite{lutz-korpa02} or even the repulsion in \cite{Kaiser:2005tu}. 
The only reliable evidence, obtained from kaonic atoms, is that the $\Sigma(1195)$ requires attraction at small densities. The $\Sigma^*(1385)$ stays close to the free position for both approaches, in contrast to \cite{lutz-korpa02}, and increases the width from 30 MeV in free space to 80 MeV when pions are dressed.

The self-energy of $\bar K$ at $\rho_0$ as function of energy is shown in Fig.~\ref{fig:3} (left) for two momenta. We show the $s$-wave component only dressing kaons (dotted lines), dressing also pions (dashed lines) and the $p$-wave contributions for the latest approach (solid lines).  We observe a small imaginary part at subthreshold for zero momentum coming from  $\bar K N N \to \Sigma N$, $\Lambda N$. The small $p$-wave strength is due the Fermi motion of nucleons which  produces a slight repulsion in the real part of the self-energy since the energies that come into play are above the
 $\Lambda$, $\Sigma$ and $\Sigma^*$ excitations. At finite momentum of 450 MeV/c, the imaginary part shows the $\Sigma N^{-1}$ component at 300 MeV and the $\Sigma^* N^{-1}$ one around 550 MeV. In this case, the $p$-wave components determine the imaginary part. The optical potential calculated from the full self-energy as $\Pi_k(q,\varepsilon(q))/(2\varepsilon(q))$, where $\varepsilon(q)$ is the quasiparticle energy $\varepsilon(q)^2=m_{\bar K}^2+q^2+{\rm Re}\Pi(q,\varepsilon(q))$, changes from -40 MeV at $\rho_0/2$ to -70 MeV at $2 \rho_0$ for $\vec{q}=0$ when only $\bar K$ are dressed self-consistently. A similar shift from -30 MeV to -80 MeV in the range of 
$\rho_0/2$ to $2 \rho_0$ is obtained when pions are also dressed. Moreover, the imaginary part is sizeable for both approaches, in agreement with the self-energy.

Finally, the antikaon spectral function at zero momentum for different densities and the two approaches is shown in Fig.~\ref{fig:3} (right). The spectral function does not show a Breit-Wigner behaviour. The slow fall off on the right-hand side of the quasiparticle peak is due to  $\Lambda(1405)N^{-1}$ excitation and the $p$-wave components are the result of the Fermi motion of nucleons. With increasing density, the quasiparticle peak gains attraction and the spectral function is diluted. The small peak observed on the right-hand side of the quasiparticle peak at $2\rho_0$ is due to $\Sigma^*(1385)N^{-1}$.

In summary, we have investigated the properties of the $s$- and $p$-wave $\bar K$ self-energy in nuclear matter within the context of a chiral unitary approach. We have shown that the on-shell factorization of the amplitudes cannot be applied for the in-medium corrections to the $p$-wave amplitudes. Furthermore, we have studied the properties of the $\Lambda$, $\Sigma$ and $\Sigma^*$ hyperons in nuclear matter. While $\Lambda$ and  $\Sigma$ feel an attractive potential of -30 MeV at $\rho_0$, the $\Sigma^*$ barely changes its mass but develops a width of 80 MeV. The contribution of the $p$-wave components to the $\bar K$ self-energy and, hence, to the spectral function  are small for low-momenta but considerable at subthreshold energies for finite momenta due to $\bar K N \rightarrow \Sigma$ conversion. The antikaon potential obtained from the self-energy can only give $\bar K$ states bound by no more than 50 MeV but with a sizeable width of 100 MeV. Therefore, deep and bound $\bar K$ states are not expected, in agreement with previous self-consistent calculations.

\vspace{-0.6cm}



%

\end{document}